# THE NATURE OF THERMAL BLACK BODY RADIATION

Fedor V.Prigara

Institute of Microelectronics and Informatics, Russian Academy of Sciences, 21 Universitetskaya, 150007 Yaroslavl, Russia; fprigara@rambler.ru

## ABSTRACT

It was shown recently that thermal radio emission has a stimulated character, and it is quite possible that thermal black body radiation in other spectral ranges also has an induced origin. The induced origin of thermal black body emission leads to important astrophysical consequences, such as the existence of laser type sources and thermal harmonics in stellar spectra.

*Subject headings:* radiation mechanisms: thermal

## 1. MASER AMPLIFICATION OF THERMAL RADIO EMISSION

The induced origin of thermal radio emission follows from the relations between Einstein's coefficients for a spontaneous and induced emission of radiation (Prigara 2001a). The strong argument in a favour of an induced character of thermal black body emission is that the spectral energy density in the whole range of spectrum is described by a single Planck's function. So if thermal radio emission is stimulated, then thermal emission in other spectral regions also should have the induced character.

The existence of maser sources associated with gas nebulae and galactic nuclei (Miyoshi et al. 1995) is closely connected with the stimulated origin of thermal radio emission. The high brightness temperatures of compact, flat-spectrum radio sources (Bower & Backer 1998; Nagar, Wilson, & Falcke 2001; Ulvestad & Ho 2001) may be explained by a maser amplification of thermal radio emission. A maser mechanism of emission is supported by a rapid variability of total and polarized flux density on timescales less than 2 months (Bower et al. 2001). Such a variability is characteristic for non-saturated maser sources. Note that the spherical accretion models with the synchrotron mechanism of emission are unable to explain the flat or slightly inverted spectra of low-luminosity active galactic nuclei (Nagar et al. 2001; Ulvestad & Ho 2001). The Blandford and Konigl theory used by Nagar et al. (2001) is in some respects similar to the gaseous disk model (Prigara 2001b), the latter being more simple and free from indefinite parameters, such as an empirical spectral index.

It is shown by Siodmiak & Tylenda (2001) that the standard theory of thermal radio emission which does not take into account the induced character of emission cannot explain the radio spectra of planetary nebulae at high frequencies without an introduction of indefinite parameters.

## 2. LASER TYPE SOURCES AND THERMAL HARMONICS

Assuming the stimulated character of thermal black body radiation, one may expect that the laser type sources in other spectral regions also can exist. One possible example of a laser source is a gamma-laser in the Galactic Centre emitting in the line 0.511 MeV (Lingenfelter & Ramaty 1982). Another effect presumably related with non-saturated laser sources is the large fluctuations in widths of solar EUV lines (Athay & White 1980). The model of the Sun as an aggregate of small cells with random phases (Athay & White 1980) is analogous to an aggregate of molecular resonators used by Prigara (2001b).

A possible X-ray laser emitting in the Fe K$\alpha$ line at 6.49 keV was recently discovered in the radio-loud quasar MG J0414+0534 (Chartas et al. 2001). A five-fold increase of the equivalent width of the Fe K$\alpha$ line was detected. The continuum emission component did not follow this sudden enhancement of the iron line.

The polarization of optical emission lines in the spectra of radio galaxies (Tadhunter et al. 2001) and of rapidly varying continuum in the spectra of BL Lacertae objects (Rector & Stocke 2001) also may be explained by a laser mechanism of emission. Rapid variations of [OIII] line profiles in Seyfert galaxies is similar effect.

A molecular resonator (Prigara 2001a) can emit second and higher harmonics. Such thermal harmonics are likely to be observed in the infrared spectrum of planetary nebula NGC 7027 (Pottasch 1984). There is a clear indication of second thermal harmonic in the optical spectrum of the Sun. This effect may determine, partially or in whole, the deviation of the spectral energy distribution of the Sun from

Planck's function. It is quite possible that thermal harmonics play a significant role in spectral differences between main-sequence stars.

To summarise, there are strong theoretical and observational indications of the stimulated origin of thermal black body radiation.